\documentclass[onecollarge]{svjour2}       
\smartqed  
\usepackage{graphicx}
\usepackage{epstopdf}
\usepackage{color}
\usepackage{latexsym}
\usepackage{amssymb}
\usepackage{textcomp}

\newcommand{\bra}[1]{\langle #1 |}
\newcommand{\ket}[1]{| #1 \rangle}

\newcommand{\be}{\begin{equation}}
\newcommand{\ee}{\end{equation}}
\newcommand{\ba}{\begin{eqnarray}}
\newcommand{\ea}{\end{eqnarray}}

\newcommand{\ematriz}[3]{\left\langle {#1} \left|{#2}\right|{#3}\right\rangle}	
\newcommand{\proj}[2]{\left| {#1} \right\rangle\!\left\langle {#2} \right|}

\begin{document}

\title{Quantum Thermometry}


\author{\large Robert B. Mann \and Eduardo Mart\'in-Mart\'inez}
\institute{ 
           R. B. Mann \at Department of Physics and Astronomy  \at Institute for Quantum Computing, University of Waterloo, 
           Waterloo, ON, Canada
              \email{rbmann@uwaterloo.ca} 
    \and       
           E. Mart\'{\i}n-Mart\'{\i}nez \at Department of Applied Mathematics \at Institute for Quantum Computing, University of Waterloo
Waterloo, ON, Canada
             \email{emartinm@uwaterloo.ca}
            }

\date{ }

\maketitle

\begin{abstract}
In this review article we revisit and spell out the details of previous work on how Berry phase can be used to construct a precision quantum thermometer. An important advantage of such a scheme is that there is no need for the thermometer to acquire thermal equilibrium with the sample. This reduces measurement times and avoids precision limitations. We also review how such methods can be used to detect the Unruh effect.
\end{abstract}
 
 \section{Introduction}
\label{intro}

{\it Introduction.--}
The science of thermometry is nearly 400 years old, dating back to the work of Galileo, Biancani, Sagredo, and Fludd. It was Ferdinando II de Medici, who constructed the first genuine thermometer, which consisted of sealed tubes  with a bulb and stem that were partially filled with alcohol.  This device was independent of air pressure, and so the expansion of the liquid within depended only on the temperature of its surrounding environment.

Standard commercially available thermometers are precise to approximately $0.01$ Celsius degrees.  Precision can be considerably improved by using resistance temperature detectors (RTDs).  These devices exploit the fact that electrical resistance of platinum responds to  temperature in a precisely known way: by  0.00385 ohm per ohm of resistance for each degree C of temperature change.
The best such instruments available can resolve temperature differences as small as $10^{-7}$ degrees C.  
However all such devices require a given substance (alcohol,mercury, platinum) to come into thermal equilibrium with its environment.

Here we report on a class of proposed thermometers that make use of quantum effects to determine temperature. These devices make use of temp\-erature-sensitive quantum effects to yield information about temperature. They do not need to come into thermal equilibrium with an environment.  Furthermore, they are capable of considerable precision, of order $10^{-6}$ degrees C.

The primary quantum-mechanical effect we exploit is the geometric phase. Upon interacting with a quantum field, the state of a point-like quantum system with discrete energy levels (e.g. an atom) acquires a geometric phase \cite{BerryOriginal} that is dependent on the state of the field.  If the field is in a thermal state, this geometric phase encodes information about its temperature \cite{2011arXiv1112.3530M}.

Our work was motivated by a recent proposal for measuring the Unruh effect  at low accelerations that also exploits the geometric phase \cite{prl}.  For several decades now it has been known that a (not quite) straightforward application of quantum field theory in relativistic settings  implies that  the vacuum state of a quantum field corresponds to a thermal state when described by uniformly accelerated observers, a phenomenon known as the Unruh effect \cite{Unruh0,bigreview}.  Direct detection has continued to elude empirical scrutiny since the associated temperature of the thermal bath (the Unruh temperature) is smaller than a $1^o$ Kelvin, even for accelerations
as high as $10^{21}${m}/{s}$^2$.  With current technology, sustained accelerations higher than $10^{26}${m}/{s}$^2$ are required to detect the effect \cite{ChenTaj,bigreview}, one of the  main experimental goals of our time \cite{experiments}.  Observation of this phenomenon would settle long standing controversies  concerning the  very existence of the effect, and would also 
provide strong indirect empirical support  for  black hole evaporation and radiation \cite{Hawking}.  Detection of the Unruh effect would have an immediate impact in many fields such as astrophysics \cite{Astronature}, cosmology \cite{Cosmo}, black hole physics \cite{Bholes}, particle physicsÊ\cite{Base},  quantum gravity \cite{Qg} and relativistic quantum information \cite{Alicefalls}.  
A number of proposals have been put forward to this end.  These include  analog systems such as fluids \cite{Unruhan,Weinfurtner:2010nu}, Bose-Einstein condensates \cite{garay}, optical fibers \cite{optfib}, slow light \cite{slowlight},  superconducting circuits \cite{supercond} and trapped ions \cite{ions}.  The best case scenarios yield Unruh temperatures  of the order of nanokelvin that remain very difficult to detect.

\begin{figure}[h]
\begin{center}
\includegraphics[width=.50\textwidth]{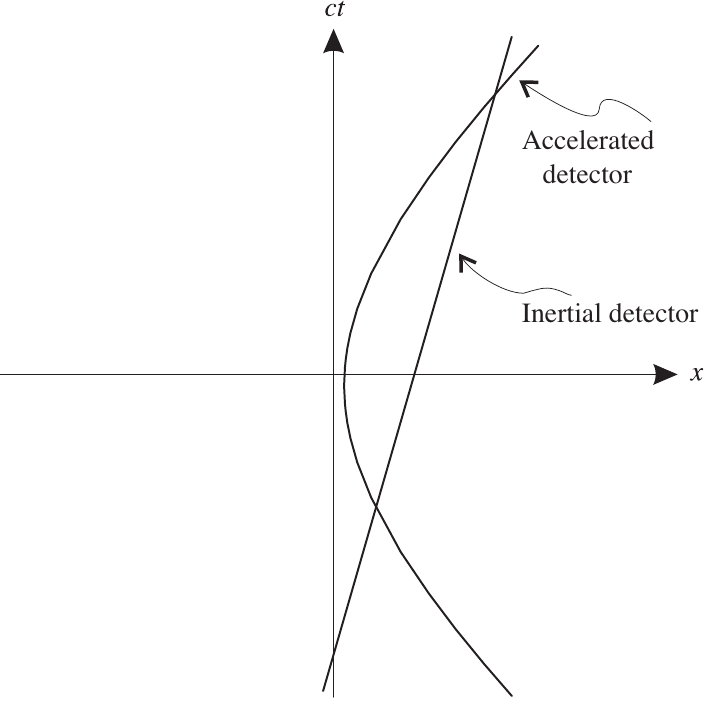}
\caption{Trajectories for an inertial and accelerated detector.}
\label{setup}
\end{center}\label{trajs}
\end{figure}

The key feature we exploit here is the conjunction of the geometric phase with methods from (relativistic) quantum information. The former effect, first noticed by Berry, is that when the parameters of the Hamiltonian of a quantum system are varied in a cyclic and adiabatic fashion,
its  eigenstates acquire a phase  (in addition to the usual dynamical phase) \cite{BerryOriginal}.  The latter effect, noted more recently \cite{Alicefalls}, is that acceleration degrades quantum entanglement.  The conjunction of these two notions suggests that a point-like detector interacting with a quantum field can acquire a geometric phase due to its movement in space-time under certain conditions, inertial detectors acquire a phase different from that of accelerated ones (Fig. \ref{trajs}).  Hence the phase encodes information about the Unruh temperature; as we shall see accelerations as low as $10^{17}$ {m}/{s}$^2$ can be detected in this manner.  More generally, a detector moving through
a thermal bath at one temperature will acquire a phase different from movement through a bath at another temperature, yielding a form of thermometry dependent only on quantum effects.

At this point our methods are best suited for measuring the temperature of radiation confined to a cavity.
The thermometer is a 2-level atomic system that interacts with the radiation via a Hamiltonian whose properties we consider in the next section.  We shall model the radiation as a massless scalar field and the 2-level system as an Unruh-de Witt detector.    While this approach is restrictive, its applications are conceivably quite broad. For example the temperature of a gas could be measured by putting it in thermal equilibrium with radiation, whose temperature could then be detected using our approach.

\section{Hamiltonian Diagonalization}

To illustrate how such quantum thermometry works, consider a massless scalar field in a cavity from the perspective of inertial observers moving in a flat ($1+1$)-dimensional spacetime.   The field in the cavity
is taken to be in a thermal state of a given temperature, either because it is in thermal equilibrium with an environmental reservoir, or due to the Unruh effect, in which uniformly accelerated observers will experience this state to be that of a thermal state with temperature  ${T_U=\hbar a / (2\pi c k_B)}$ where $a$ is the observer's acceleration, $c$ the speed of light and $k_B$ Boltzmann's constant (the so-called Unruh temperature). In either case, the Hamiltonian is that of a point-like harmonic-oscillator-detector endowed with an internal structure that couples linearly to the scalar field $\phi(x(t))$ at a point $x(t)$ corresponding to the world line of the detector; explicitly 
\begin{equation}\label{goodham2}
 H_T\!=\!\Omega_a a^\dagger a+\Omega_b b^\dagger b + \lambda (b+b^\dagger)[a^\dagger e^{i(kx-\Omega_a t)}+ a e^{-i(kx-\Omega_a t)}]
 \end{equation}
 where $\Omega_a$ and $\Omega_b$ are the field and atom frequencies respectively and $\lambda$ is the coupling frequency.  Here
$b^{\dagger}$ and $b$ are the ladder operators of the harmonic oscillator and the field operator takes the form $\hat\phi(x(t))\approx\hat\phi_k(x(t))\propto \left[a\, e^{i(kx-\Omega_a t)}+a^\dagger\, e^{-i(kx-\Omega_a t)}\right]$, where $a^{\dagger}$ and $a$ are creation and anihilation operators associated with the field mode $k$ with frequency $|k|=\Omega_a$. This model is a type of Unruh-DeWitt detector \cite{bigreview} which has been previously studied in \cite{LinHu}.  Note that in \ref{goodham2} we employ a mixed picture (where the detector's operators are time independent) as opposed to a Heisenberg or an interaction picture; this setting is more convenient for the calculations we wish to carry out.
 
In a realistic scenario, the oscillator couples to a peaked distribution of field modes.  As a first attempt, we will  sharpen this peak to   approach a delta function by assuming that only one mode of the field is coupled to the detector.  A cavity at finite temperature can be employed to engineer such a single mode interaction; its modes vary over a wide range of frequencies, but if one of them is close to the detector's natural frequency,   the detector effectively interacts only with this single mode.  Although the employment of such simplified models is extensive in quantum optics (for instance the Jaynes-Cummings model \cite{Scullybook}), these models are known to be problematic in very short time regimes for which the interaction of the detector with the off-resonant modes cannot be neglected, leading to the appearance of unphysical faster-than-light signalling \cite{Fay,Jonsson:2013ikb}. For the time scales of our idealized setting, the corrections coming from additional off-resonant modes do not destroy the effect  \cite{Jonsson:2013ikb,Marvy,Eric,Ivette}.

To illustrate this,  assume that the fundamental mode of a length $L$ cavity is in resonance with the energy gap of the atom. In order to neglect the contribution to the time evolution given by the interaction of the atom with the second, third and subsequent harmonics,  the energy gap between the fundamental mode of the cavity and the second harmonic must be much larger than the gap between the ground and excited states of the atom. The gap between the different harmonics in the cavity is proportional to $1/L$, so   for this single mode model   to work we require that $L$ is small enough to ensure there is a sufficiently large gap between the modes. To be sure that any of the effects described by this simplified model do not come from any spurious non-causal behaviour, we would have to make sure that the relevant times of evolution are much larger than the light-crossing time of the cavity. As we will see below, in our setting such characteristic times  are of order $\Omega^{-1}$ where $\Omega$ is the atom gap. In our scenario, for the experimentally  feasible values considered below, an atom gap of 1 Ghz yields an evolution time of 1 ns in the least  favourable scenario we consider.  The length of the cavity such that the light crossing time is precisely $t_{c}=1$ ns is $L= 0.3$ m. For time evolution scales to be much larger than the light crossing time of the cavity we need to consider a cavity of centimetres or millimetres \cite{Jonsson:2013ikb}, something very feasible from the experimental viewpoint. For the other cases proposed here the gaps as small as 1 Mhz;  ensuring non-signaling implies that the cavity should only be smaller than hundreds of meters. This requirement is obviously easy to fulfill experimentally.

The procedure for computing the geometric phase is to first diagonalize the Hamiltonian  (\ref{goodham2}). This can be done analytically and details are given in the appendix. The unitary operator that accomplishes this depends on the parameters $(u,v,s,p,\omega_a,\omega_b)$, each of which are functions of $\lambda$ and the detector frequencies $\Omega_a,\Omega_b$ -- only three of these turn out to be independent, and we take them to be $(v, \omega_a, \omega_b)$ The associated eigenstates of  (\ref{goodham2}) are $U^{\dagger}|n_f n_d\rangle$, where $|n_f n_d\rangle$ are the eigenstates of $H_0(\omega_a,\omega_b)=\omega_a\, a^\dagger a+\omega_b\, b^\dagger b$ and $U=S_a S_b  D_{ab}\hat S_b R_a$. The subscripts $f$ and $d$ respectively refer to field modes (on which the $(a,a^\dagger)$ act) and
detector modes (on which the $(b,b^\dagger)$ act). The operators
\begin{equation}
\begin{array}{ll}
\nonumber D_{ab}=\exp\big[s (a^\dagger b -  a b^\dagger)\big],&S_{a}=\exp\big[\frac12 u({a^\dagger}^2 - a^2)\big],\\*
\nonumber S_{b}=\exp\big[\frac12 v(b^2 - {b^\dagger}^2) \big],&\hat S_{b}=\exp\big[p\, ({b^\dagger}^2 - b^2)\big]
\end{array}
\end{equation}
 and $R_a=\exp\big(-i\varphi\, {a^\dagger a}\big)$ are the two-mode displacement, single-mode squeezing and phase rotation operators \cite{Scullybook}, respectively. Their action on the various creation and annihilation operators is given in the appendix.
 
The next step is to compute the geometric phase under cyclic evolution of the parameters $(v, \omega_a, \omega_b)$ for a detector
 (an inertial atom) interacting with an eigenstate of the Hamiltonian.  The third step is to   repeat this for an atom interacting with a field mode in a thermal state. Finally the (temperature-dependent) net geometric phase can be computed. 
 
\begin{figure}[h]
\begin{center}
\includegraphics[width=1.0\textwidth]{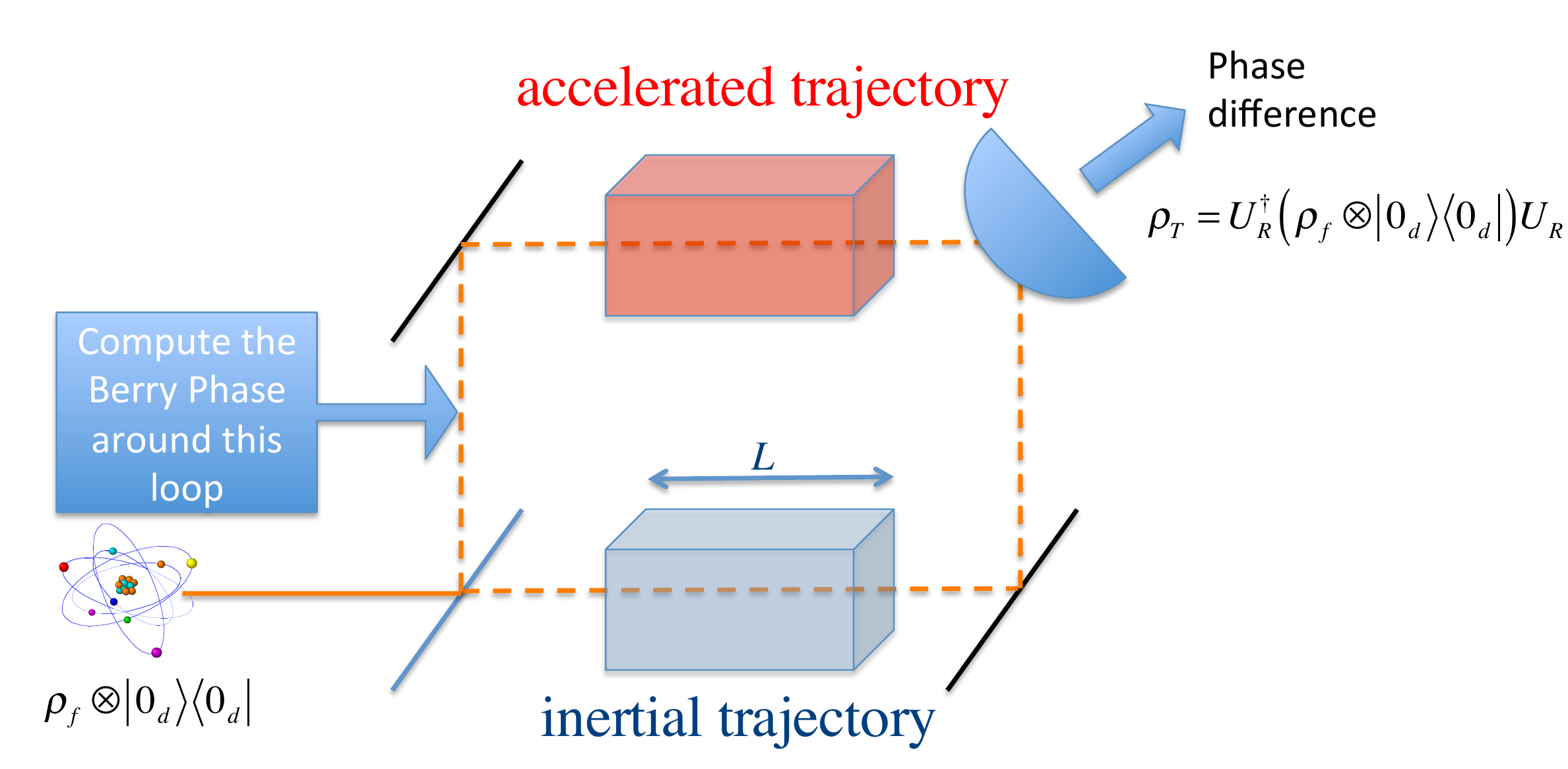}
\caption{Schematic diagram for measuring the Unruh effect via the geometric phase.  The system is initially in the mixed state  $\rho_f\otimes{\proj{0_d}{0_d}}$; upon suddenly turning on the interaction, a general state $\ket{n_f 0_d} \simeq U_R^\dagger\ket{i_f j_d}$ where $n_f=i_f+j_d$ for eigenstates $\ket{i_f,j_d}$. By making a projective measurement we can  verify that the detector is still in its ground state, ensuring that state of the joint system is  $\rho_T= U_R^\dagger \left(\rho_f\otimes{\proj{0_d}{0_d}}\right)U_R$.
 }
\label{mziberry}
\end{center}
\end{figure}

 A schematic diagram is given in figure \ref{mziberry}.
Note that it is the displacement of the detector in space-time that generates a cyclic change in the Hamiltonian, with the phase
  $\varphi=kx-\Omega_a t$, of  the field operators completing a $2\pi$ cycle in  time $\Delta t\sim\Omega_a^{-1}$, where  $(t,x)$
are Minkowski coordinates (a convenient choice for inertial observers).  

Before the interaction between the field and the detector is switched on, the field is in the vacuum state and the detector in the ground state and so the system is in the state $\ket{0_f0_d}$. We find that after the coupling is   switched on   the state of the system is 
\begin{equation}\label{adiab}
\ket{0_f0_d}=\sum_{n,m} \ematriz{n_fm_d}{U}{00}U^\dagger \ket{n_fm_d}.
\end{equation}
in the sudden  switching approximation\footnote{Suddenly switching on the coupling is known to be problematic since it can give rise to divergent results. However, in this case such problems are avoided 
because we are considering an effective $(1+1)$ dimensional setting. In $(3+1)$ dimensions these divergences can be treated by introducing a continuous switching function \cite{Louko:2007mu}; the results are qualitatively the same.}
In the coupling regimes we consider  
$$
\ematriz{n_fm_d}{U}{00}= \bra{n_fm_d}{S_aS_bD_{ab}\hat S_b R_a}\ket{00}\approx \delta_{n_f0}\delta_{m_d0}
$$  
which can be demonstrated numerically.  Hence for either cavity we have
\begin{equation}\label{adiab}
\ket{\psi_{00}}=\sum_{n,m} \ematriz{n_fm_d}{U}{00}U^\dagger \ket{n_f m_d} = U^\dagger \ket{0_f 0_d} + {\cal O}(\lambda^2)
\end{equation}
and so for small $\lambda$ all changes are adiabatic.
After the coupling is suddenly switched on and the state of the system is  $U^\dagger\ket{0_f 0_d}$, the movement of the detector in spacetime, which can be considered cyclic and adiabatic, generates a Berry phase. 

\section{Berry Phase Computation}

The Berry phase $\gamma$ acquired by the eigenstate $\ket{\psi(t)}$  of a system whose Hamiltonian depends on $k$ parameters $R_1(t),\dots,R_k(t)$ that vary cyclicly and adiabatically is given by
\begin{equation}\label{Berry}
i\gamma=\oint_R\,  \mathbf{A} \cdot  \textrm{d} \mathbf{R}
\end{equation}
where
\[\mathbf{A} =\left(\!\begin{array}{c}
\bra{\psi(t)}\partial_{R_1}\ket{\psi(t)}\\
\bra{\psi(t)}\partial_{R_2}\ket{\psi(t)}\\
 \vdots\\
 \bra{\psi(t)}\partial_{R_k}\ket{\psi(t)}
\end{array}\!\right)\]
 and $R$ is a closed trajectory in the parameter space \cite{BerryOriginal,Amj}.
 We calculate the Berry phase acquired by an eigenstate of the Hamiltonian under cyclic and adiabatic evolution of parameters  $(v,\varphi,\omega_a,\omega_b)$. 

The only relevant parameter that will vary under time evolution is  $\varphi$.  It is straightforward to see that the variation of the parameter $\nu$ will not generate a Berry phase since
\begin{eqnarray}
A_\nu &\propto& \bra{n_f m_d}S_aD_{ab}R_a\partial_{v} (R_a^\dagger D^\dagger_{ab} S_a^\dagger)\ket{n_f m_d} \nonumber\\ 
&=&\bra{n_f m_d}S_aD_{ab}\partial_{v} ( D^\dagger_{ab} S_a^\dagger)\ket{n_f m_d} \nonumber\\
&=&0.
\end{eqnarray}
Because there are no number operators inside the bra and the ket after derivation and action with all the operators, the only contribution to the Berry phase 
\begin{equation}
i\gamma_I = \oint_{\varphi\in[0,2\pi)}\!\!\!\!\!\!\!\!\!\!\!\!\!\!\!  \mathbf{A} \cdot  \textrm{d} \mathbf{R} = \int_{0}^{2\pi}\textrm{d}\varphi\, A_{\varphi}
\end{equation}
comes from the variation of the parameter $\varphi$.

Now, since $\partial_{\varphi} R_a^\dagger = i R_a^\dagger\, a^\dagger a $ and the other operators do not depend on $\varphi$ we can readily compute
\begin{eqnarray}
A_{\varphi}&=&\bra{n_f m_d}S_aS_bD_{ab}R_a\partial_{\varphi} (R_a^\dagger D^\dagger_{ab} S_b^\dagger S_a^\dagger)\ket{n_f m_d} \nonumber\\
&=&i\bra{n_f m_d}S_a S_b D_{ab}R_a\, a^\dagger a\, R_a^\dagger D^\dagger_{ab}S_b^\dagger  S_a^\dagger\ket{n_f m_d} \nonumber\\
&=&i\bra{n_f m_d}S_b S_aD_{ab}\, a^\dagger a\,  D^\dagger_{ab} S_b^\dagger S_a^\dagger\ket{n_f m_d}
\end{eqnarray}
and making use of the relation (\ref{resultdisp}) in the appendix, we know that
\begin{equation}
D(s,\phi)\,a^\dagger a\, D^\dagger(s,\phi)= a^\dagger a\, \cos^2 s + b^\dagger b\,\sin^2 s - \frac{1}{2}\sin 2s \left(a^\dagger b\, e^{i\phi}+b^\dagger a\, e^{-i\phi}\right)
\end{equation}
and that
\begin{equation} 
S_a (t,\theta)\,a^\dagger a\,S^\dagger_a(t,\theta) = a^\dagger a \cosh 2 t -\frac12 \big(a^\dagger a^\dagger e^{-i\theta} + a a\, e^{i\theta} \big)\sinh 2t+  \sinh^2 t
\end{equation}
and so we can successively commute the number operators and compute its integral over the parameter space. This yields the Berry phase 
\begin{eqnarray}
\nonumber \gamma_{I_{n_f n_d}}=&\,2\pi\bigg[\frac{\omega_a\, n_d \cosh(2v)\sinh[2(C-v)]}{\omega_a\sinh[2v]+\omega_b\sinh(2v)}\\
+&\frac{\omega_b\, n_f \sinh(2v)\cosh[2(C-v)]}{\omega_a\sinh[2(C-v)]+\omega_b\sinh(2v)}+T_{00}\bigg] 
\end{eqnarray}
acquired by an eigenstate $U^\dagger\ket{n_f n_d}$,  where 
\begin{equation}
T_{00}=\frac{\omega_a \sin^2 v \sinh [2(C-v)]+\omega_b\,\sinh(2v)\sinh^2 (C-v)}{\omega_a\, \sinh[2(C-v)]+\omega_b\,\sinh (2v)}
\end{equation}
with $C=\frac{1}{2}\ln\left(\omega_a/\omega_b\right)$ and  $\omega_a/\omega_b>e^{2v}$.

In the special case of the ground state ($n_f=n_d=0$) we obtain
\begin{eqnarray}
\gamma_{I_{00}}&=2\pi\,T_{00}.
\end{eqnarray}
Note that the ground state is non-degenerate and the  gaps between energy levels are independent of time.

\section{Thermometry from Geometric Phase}

We here discuss how to utilize  the geometric phase as a probe of thermal systems \cite{2011arXiv1112.3530M}.  The idea is to use 
an atomic interferometer as a thermometer, measuring the  temperature of a cold medium by comparison with a hotter thermal source of approximately known temperature. The geometric phase  acquired by an atom  interacting with a thermal state instead of an eigenstate of the Hamiltonian (see \cite{Vlatko}) provides a measure of the temperature of the thermal state without any requirement that the atom comes into thermal equilibrium with the colder source.

Consider  a bosonic medium  at temperature $T$ contained within a cavity.  The density matrix is
\[\rho_{T}=\bigotimes_{\omega}\frac{1}{\cosh^2 r_{_T} }\sum \tanh^{2n} r_{_T} \proj{n_\omega}{n_\omega}\]
where
\[\tanh r_{_T} = \exp\left(-\frac{\hbar  \omega}{2k_BT}\right).\]
The initial state of  the field and atom the system is  the mixed state  $\proj{0}{0}\otimes\rho_{T}$. Upon adiabatically turning
on the interaction, the system evolves into the state  $\rho= U^\dagger \left({\proj{0}{0}}\otimes\rho_f\right)U$.

The mixed state $\rho$ acquires a geometric phase $\gamma = {\textrm Re}(\eta)$, with  
\begin{equation}\label{mixed}
e^{i\eta} = \sum_i\omega_ie^{i\gamma_i},
\end{equation}
 after a   a cycle of adiabatic evolution \cite{Vlatko},  
where $\gamma_i$ is the geometric phase acquired by the eigenstate $\ket{i}$. Under one cycle of evolution
 for the state $\rho_T$ we obtain
\begin{equation}
 e^{i\eta} = \frac{1}{\cosh^2 r_{_T}}\sum_{n} \tanh^{2n}r_{_T}\, e^{i\gamma_{I_{n}}} = 
 \frac{e^{i\gamma_{I_{0}}}}{\cosh^2 r_{_T}-e^{2\pi\,i G}\sinh^2 r_{_T}}, 
\end{equation}
where 
\begin{equation}
G= \frac{\nu_d\,  \sinh(2v)\cosh[2(C-v)]}{\nu_f\sinh[2(C-v)]+\nu_d\sinh(2v)}.
\end{equation}
and the parameters $(C,v,\nu_d,\nu_f)$ are as before, 
yielding
\begin{equation}
\gamma_{_T}= {\textrm  Re}\eta=\gamma_{I_{0}}- {\mathrm{Arg}}\left(\cosh^2 r_{_T} - e^{2\pi\,i G } \sinh^2 r_{_T}\right)
\end{equation}
for the acquired geometric phase.  For two  thermal environments at different temperatures the phase difference is
\begin{equation}\label{deltai}
\delta = \gamma_{_{T_1}} - \gamma_{_{T_2}}
= {\mathrm{Arg}}\left(1- e^{-\frac{\hbar\omega}{k_B T_1}-2\pi\,i G } \right)-{\mathrm{Arg}}\left(1 - e^{-\frac{\hbar\omega}{k_B T_2}-2\pi\,i G } \right)
\end{equation}

\begin{figure}[h]
\begin{center}
\includegraphics[width=.70\textwidth]{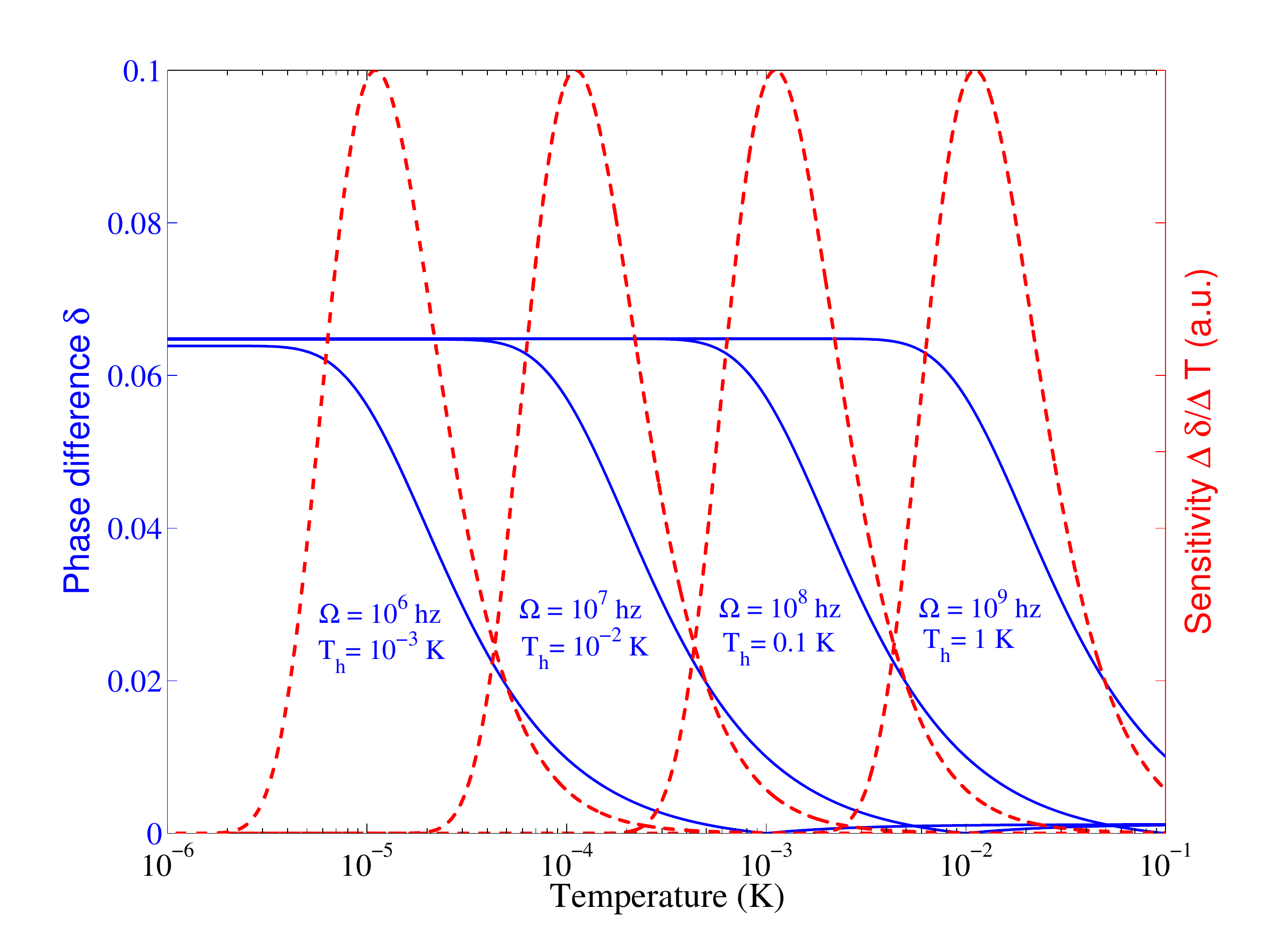}
\caption{ The geometric  phase difference $\delta$ between  2 detectors interacting with a cold and hot source of temperatures $T_{ \textrm{c}}$ and $T_{ \textrm{h}}$ respectively, as a function of the cold source temperature for different values of the atom gap and   hot source temperature. From left to right: $\Omega=10^6$ hz, $T_{ \textrm{h}}=1$ mK; $\Omega=10^7$ hz, $T_{ \textrm{h}}=10$ mK; $\Omega=10^8$ hz, $T_{ \textrm{h}}=0.1$ K; $\Omega=10^9$ hz, $T_{ \textrm{h}}=1$ K. Coupling frequency: 1.2 Khz for all the cases. Red dashed lines are sensitivity curves for all the cases previously considered.}
\label{realf1}
\end{center}
\end{figure}
\begin{figure}[h]
\begin{center}
\includegraphics[width=.80\textwidth]{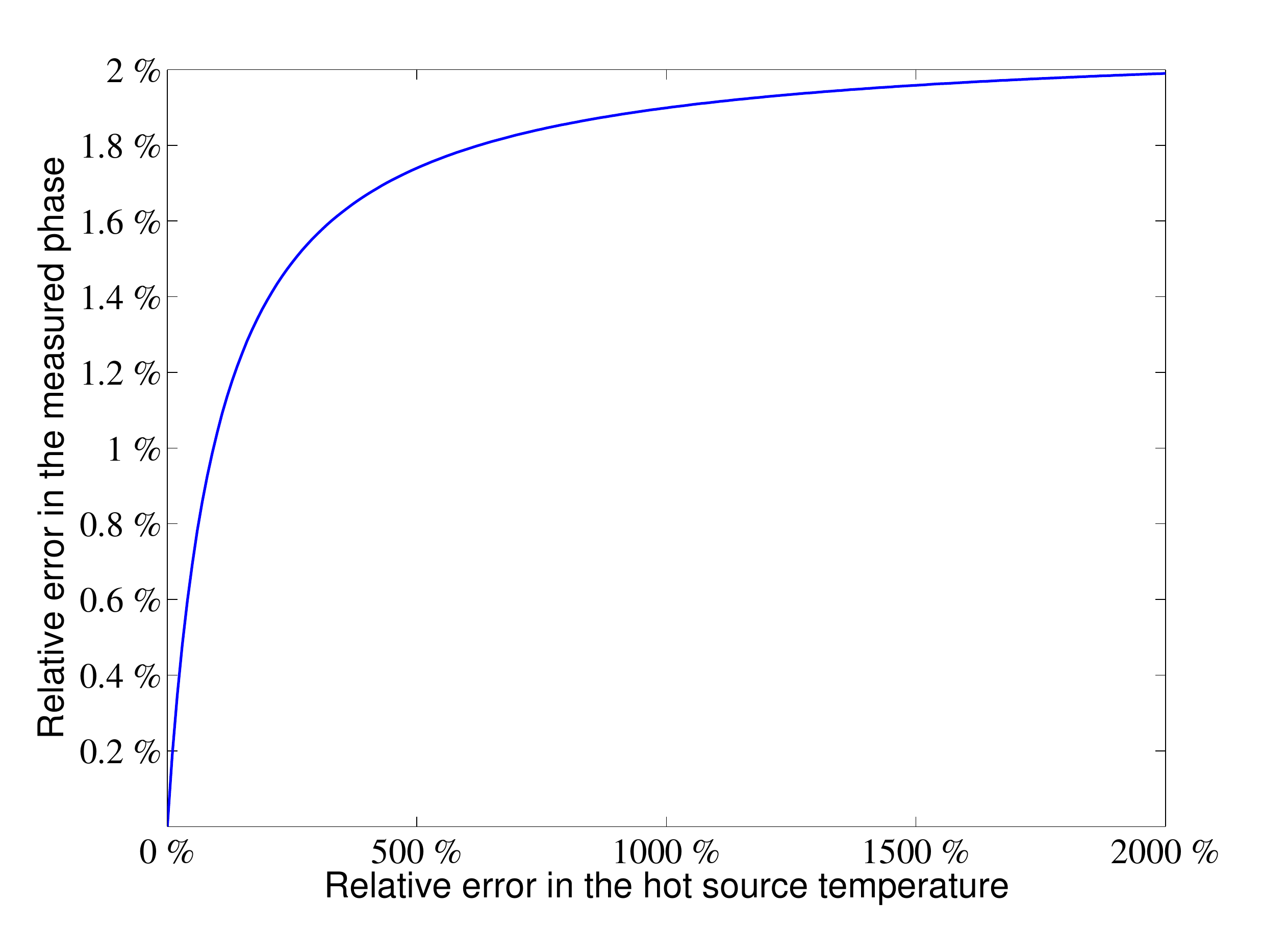}
\caption{Relative error in the Berry phase (and therefore, the determination of the temperature for the cold source) as a function of the relative error in determining the temperature of the hot source. As we see, the setting is very robust: huge changes of temperature of the hot source translate into small changes in the phase $\delta$.}
\label{precision}
\end{center}
\end{figure}

This phase difference can be quite large for realistic coupling values for atoms in cavities, as illustrated   in
fig.~\ref{realf1}. Depending on the atomic gap,  it is also very sensitive to a particular range of temperatures.  We can thus tune
the phase  $\delta$  to a particular  temperature range; we find that $\delta$ is quite sensitive to variations of the cold source but
rather insensitive to changes in the hot source. This is shown in fig. \ref{precision}:   large variations in the hot source temperature
 translate into very small variations of the measured phase, providing us with a high precision thermometer.  Furthermore, there
 is no need for the atomic (or multi-level system) probe to come into equilibrium with its thermal environment(s). 

\section{Detection of the Unruh Temperature}

Our approach for constructing a large, high-precision thermometer using atomic interferometry techniques can be exploited to provide a new test of the Unruh effect.  In this section we outline how this can be carried out.

A convenient choice of  reference frame for computing the Berry phase in the case of an accelerating atom is to use Rindler coordinates $(\tau,\xi)$, for which $\varphi\!=\!|\Omega_a|\xi-\Omega_a\tau$. The evolution is cyclic after a time $\Delta \tau = \Omega_a^{-1}$.  Adiabaticity can also be ensured  in this case since the probability of excitation is negligible for the accelerations we  consider \cite{Scully,bigreview}.  
 Although $H_T$ in (\ref{goodham2}) has the same form for both inertial and accelerated detectors, 
 in the inertial case $a,a^\dagger$ are Minkowski operators, whereas for the accelerated detector  they correspond to Rindler operators. 
 
 To make this distinction clear,  denote the respective Minkowski and Rindler operators by $U^\dagger_\textrm{M}$ and $U^\dagger_\textrm{R}$.  The state of the field is not pure for accelerated observers but rather is mixed, a key distinction from  the inertial case.   In the basis of an accelerated observer, the state $\proj{0_f}{0_f}$ transforms  to  the thermal Unruh state $\rho_f$ \cite{Unruh0,Alicefalls}, and so   before the field-detector interaction is turned on, the system is in the mixed state  $\rho_f\otimes{\proj{0_d}{0_d}}$. 
 Upon suddenly switching on the   interaction, a general state $\ket{n_f 0_d}$ evolves,  very close to a superposition of eigenstates  $U_R^\dagger\ket{i_f j_d}$ where $N_f=i_f+j_d$ in the small $\lambda$ coupling regime.  We can ensure that the state of the joint system is  $\rho_T= U_R^\dagger \left(\rho_f\otimes{\proj{0_d}{0_d}}\right)U_R$ if we verify that the detector is still in its ground state (by making a projective measurement)   immediately after switching on the interaction. 
 
  Calculating the mixed state Berry phase \cite{Vlatko} we find
\[\gamma_a=\gamma_{I}-\textrm{Arg}\left(\cosh^2 q - e^{2\pi\,i G } \sinh^2 q\right)\]
where
$\gamma_{I}$ is the inertial Berry phase, $q=\arctan\big(e^{-\pi\Omega_a c/a}\big)$ and 
\[G= \frac{\omega_b\,  \sinh(2v)\cosh[2(C-v)]}{\omega_a\sinh[2(C-v)]+\omega_b\sinh(2v)}\]
 depends  on the detector parameters.
 
\begin{figure}
\begin{center}
\includegraphics[width=.9\textwidth]{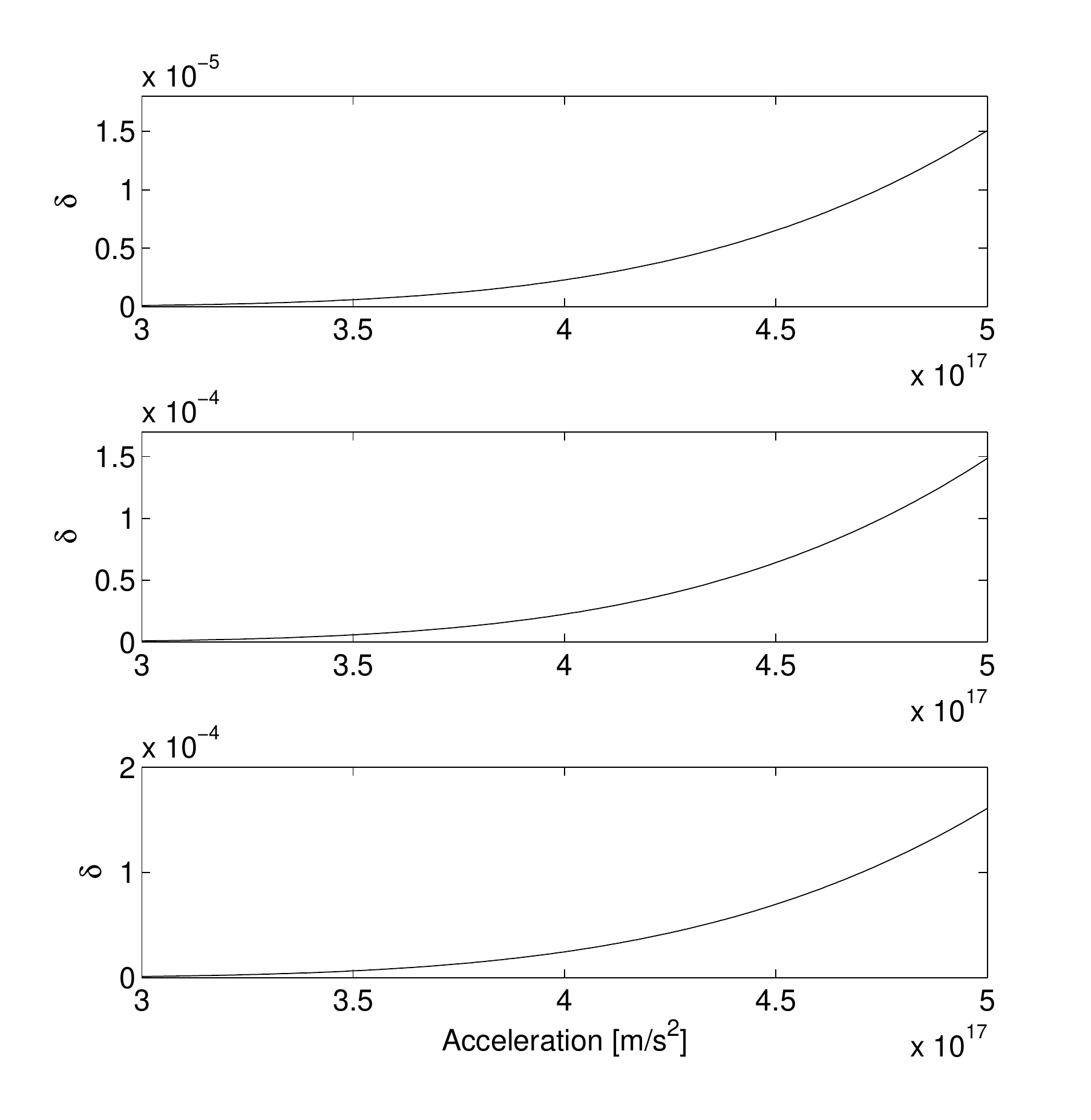}
\caption{$\delta$ for each cycle as a function of the acceleration for three different scenarios. First scenario (top): $\Omega_a\simeq 2.0$ GHz  $\Omega_b\simeq 2.0$ GHz $\lambda\simeq$ 34  Hz.
Second scenario (middle): $\Omega_a\simeq 2.0$ GHz  $\Omega_b\simeq 2.0$ GHz  $\lambda\simeq$  0.10  KHz.
Third scenario (bottom): $\Omega_a\simeq 2.0$ GHz  $\Omega_b\simeq 2.0$ GHz  $\lambda\simeq$ 0.25  KHz.}
\label{deph}
\end{center}
\end{figure}

We now compare the Berry phase acquired by the detector in the inertial and accelerated cases.
After a complete cycle in the parameter space (with a proper time $\Omega_a^{-1}$) the phase difference between an inertial and an accelerated detector is
$\delta =\gamma_{I}-  \gamma_a$.   The results are illustrated in  figure \ref{deph}, which plots the phase difference $\delta$ as a function of the acceleration for physically
relevant atomic transition frequencies \cite{revat,Scully} coupled to the electromagnetic field (in resonance with the field mode they are coupled to) in  the microwave regime ($2.0$ GHz).  We consider three different coupling strengths: 1) $\lambda\simeq$ $34$  Hz,  2) $\lambda\simeq$  $0.10$   KHz,  3)   $\lambda\simeq$ $0.25$  KHz.  

The phase difference is large enough to be detected after a single cycle (about 3.1 ns).  Evolving the system through more cycles will
enhance the phase, since the effect is cumulative.  The maximal phase difference achievable is $\delta=\pi$, corresponding to destructive
interference.  This can occur after $30000$ cycles  ( 95 $\mu$s) for an acceleration  of $a\approx 4.5\cdot10^{17}$ {m}/{s}$^2$.
For this magnitude of acceleration the atom will acquire a speed of $\approx 0.15c$ after a time $t\approx \Omega_a^{-1}$. 
 Consequently the geometric phase acquired by the joint field/atom (more generally field/detector) state
 can be used as a tool to probe the Unruh effect for accelerations as small as $10^{17}${m}/{s}$^2$.  

\section{Closing Remarks}

There are several experimental challenges to be overcome in implementing quantum thermometry.    The basic setup would be that of an interferometric experiment, as illustrated in figure \ref{mziberry}.

\begin{figure}[h]
\begin{center}
\includegraphics[width=.80\textwidth]{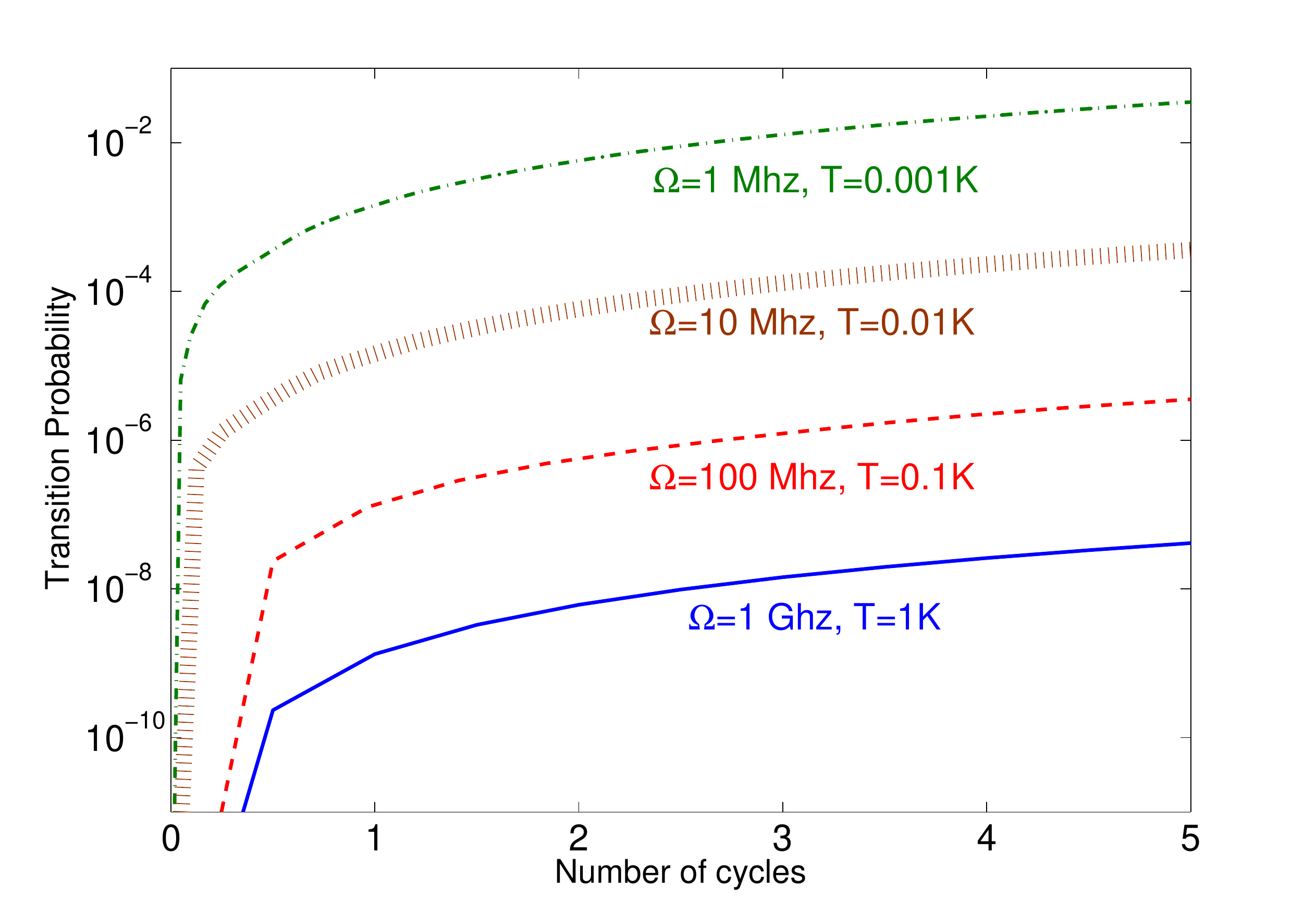}
\caption{Probability of atomic excitation in a time of a small number of cycles (1 cycle means $t= 2\pi/\omega$ s). For the 1Ghz case $P<10^{-9}$, for the Mhz case $P<10^{-3}$}
\label{appcheck}
\end{center}
\end{figure}

For (inertial) quantum thermometry in general \cite{2011arXiv1112.3530M}, it is necessary for weak adiabaticity to hold:
 there must be a near negligible  probability of finding the atom in an excited state after one cycle of evolution. Furthermore, this means that the atom does not have time to thermalize,    and the hypotheses necessary to apply Berry's formalism hold \cite{prl}. 
This requirement could fail for small atomic gaps, strong couplings, or high temperatures.  In the latter case coherence loss will occur only for thermal sources at temperatures   several orders of magnitude above the ones we are considering. When weak adiabaticity holds, the interaction time of the multi-level atom  with the thermal state must be short enough so that the only change that atomic state acquires only
a  global phase (dynamical + geometrical).  By solving the Schwinger equation (in the interaction picture)
\begin{equation}
\frac{d}{dt}\rho=-i\left[H_\mathrm{I},\rho\right]
\end{equation}
 numerically we find that the probability of finding the atom in the excited state cannot be distinguished from thermal noise  for realistic values of the coupling $\lambda$ after a short time $\sim 10^{4}\cdot 2\pi \Omega^{-1}$.  Since in our scenario the atoms interact  with the thermal bath only for very short times (1 cycle of evolution $t\approx 2\pi \Omega^{-1}$),  weak adiabaticity holds, as illustrated in fig. \ref{appcheck}.
 Even in the worst case scenario (1 Mhz gap and 1 mK temperature) the probability of excitation is $P\approx 10^{-3}\ll 1$, and values of
   $P\approx 10^{-9}\ll 1$  are conceivable.

The quantum thermometer we propose has a rather sensitive target temperature range, typically about 3 orders of magnitude as shown
in fig.  \ref{realf1}. The reference source temperature needs to be about 3 orders of magnitude larger (or smaller) than the target temperature,
though hotter reference sources are preferred since they are easier to control. 

To observe the Unruh effect, even though accelerations of $10^{17}$m/s  are nine orders of magnitude  smaller than other proposals \cite{ChenTaj}, they are still formid\-ably large, necessitating a compromise between the desired phase difference and feasibility of handling relativistic atoms.  
 Since the  phase accumulates independently of the sign of the acceleration,  alternating periods of positive and negative acceleration could perhaps be exploited  to reduce the atom's final speed, and
cancelling to some extent  the dynamical phase difference between the paths in certain settings.  For example, with current length metrology technology\footnote{$\textrm{Laserscale}^{{\textregistered}}$, {\scriptsize
http://www.gebotech.de/pdf/LaserscaleGeneralCatalog\_en\_2010\_04.pdf}} the relative dynamical phase could be controlled with a precision $\Delta \phi \approx 10^{-8}$, several orders of magnitude smaller than the Berry phase acquired in one cycle.   Recent work inspired by our quantum thermometry approach
has shown that  it is possible to take both geometric and dynamical phases into account  to build interferometric settings that are as precise as those we consider here \cite{Marvy}; no single-mode approximation is required.

Our approach can also be applied to  
Quantum Non-Demolition (QND) measurements.   It can be shown that
an atomic probe, on resonance with the target field mode we want to measure, can be sent through a cavity in a manner that does
not alter the state of light in cavity whilst acquiring  a non-negligible (and measurable) phase \cite{Marvy}.  Known as `mode invisibility', this technique allows for the effective distinction of Fock states containing very few photons via  an
interferometry setup similar to figure 2, in which one cavity contains a known state of light and the other one contains 
the unknown state of light that we want to probe.  This method
can be extended to coherent states of light that are experimentally much more controllable and easier to prepare than Fock states,
yielding information about some features of the Wigner function (such as the relative difference in the phase of a squeezing and a phase space displacement) \cite{Onuma-Kalu:2014gsa}.

Quantum thermometry, while challenging, is at the edge of experimental feasibility.  It opens up new ways to detect the Unruh effect and
perhaps to probe other phenomena (for example Bose-Einstein condensates \cite{Ivette-2}) associated with relativistic quantum information.   More generally, it can perhaps be used to probe a
variety of field/atom systems that sensitively depend on one (or more) parameters.  Work on these issues is in progress.

\begin{acknowledgements} We thank Andrzej Dragan and Ivette Fuentes for helpful comments and remarks. This work was supported in part by the Natural Sciences and Engineering Research Council of Canada. R.B.M. is grateful to Fabio Scardigli and the organizers of the Horizons of Quantum Physics conference for their invitation to speak at this meeting.  E. M-M. gratefully acknowledges the funding of the Banting Postdoctoral Fellowship Programme.
\end{acknowledgements}

\appendix

\section{Diagonalization of the Hamiltonian}

Consider a point-like detector, endowed with an internal structure, which couples linearly to a scalar field $\phi(x(t))$ at a point $x(t)$ corresponding to the world-line of the detector. The interaction Hamiltonian is of the form $H_I\propto  \hat X \hat \phi(x(t))$ where 
we have chosen the detector to be modeled by a  harmonic oscillator with frequency $\Omega_b$. In this case the operator $\hat X\propto (b^\dagger + b)$ corresponds to the detector's position where $b^{\dagger}$ and $b$ are creation and anihilation operators .
 
 Suppose that the detector couples only to a single mode of the field with frequency  $|k|=\Omega_a$. 
The field operator takes the form 
$$
\hat\phi(x(t))\approx\hat\phi_k(x(t))\propto \left[a\, e^{i(kx-\Omega_a t)}+a^\dagger\, e^{-i(kx-\Omega_a t)}\right]
$$
where $a^{\dagger}$ and $a$ are creation and annihilation operators associated with the field mode $k$.  The Hamiltonian  is therefore given by eq. (\ref{goodham2}), which is
 \begin{equation}\label{hamo}
 H_T\!=\!\Omega_a a^\dagger a+\Omega_b b^\dagger b + \lambda (b+b^\dagger)[a^\dagger e^{i(kx-\Omega_a t)}+ a e^{-i(kx-\Omega_a t)}].
 \end{equation}
 where $\lambda$ is the coupling frequency, and  resembles an Unruh-DeWitt detector in the case where the atom interacts with a single mode of the field.   In what follows we employ a mixed picture, in which the detector's operators are time independent,
in contrast to standard approaches that   employ the interaction picture. The latter is the most convenient picture for computing transition probabilities, whereas we find the former  mathematically more convenient for Berry phase calculations.
 
To diagonalize the  Hamiltonian (\ref{hamo}) we begin with a diagonal Hamiltonian of the form
\begin{equation}\label{origen}
H_0={\omega_a} a^\dagger a + {\omega_b} b^\dagger b
\end{equation}
Our objective is to obtain the unitary transformation that diagonalises (\ref{hamo}). We shall do this by 
finding the unitary transformation that transforms the Hamiltonian (\ref{origen}) into (\ref{hamo}); the inverse operator is then the operator that diagonalizes (\ref{hamo}).    Once we obtain its eigenstates and eigenvalues we will be able to compute the geometrical phase acquired after cyclic evolution. Throughout we shall make use of the relation 
\begin{equation}
e^{B}A\, e^{-B} = \exp\left(ad_B \right) A =  A + [B,A] +  \frac{1}{2}[B,[B,A]] + \cdots \nonumber
\end{equation}
where $ad_B(A) \equiv [B,A] $.

Let us introduce the single mode squeeze operator  
\[S_{a}=\exp\left(\alpha^* {a^\dagger}^2 - \alpha a^2\right)\]
whose  action on the creation/annihilation operators
\begin{eqnarray}\label{a1modesqeez}
\nonumber S^\dagger_{a}\,a\, S_{a}&=&  a\, \cosh t + a^\dagger e^{-i\theta}\sinh t\\*
S^\dagger_{a}a^\dagger S_{a}&=& a^\dagger\, \cosh t + a e^{i\theta}\sinh t
\end{eqnarray}
is straightforward to show upon setting $\alpha= \frac{t}{2} e^{i\theta}$. 

We first apply a 2 single mode squeeze to the Hamiltonian $H_0$ via
$$
H_{1s}=S^\dagger_a(u,\theta_a) S^\dagger_b(v,\theta_b) H_0 S_b(v,\theta_b)S_a(u,\theta_a)
$$
obtaining
\begin{eqnarray}\label{1step}
\nonumber H_{1s}&=&{\omega_a} \left[a^\dagger a\, \cosh 2u +\frac12\sinh 2u \left(a^\dagger a^\dagger e^{-i\theta_a} + a a\, e^{i\theta_a} \right)\right]\\
&& \quad +{\omega_b} \left[b^\dagger b\cosh 2 v  +\frac12\sinh 2v \left(b^\dagger b^\dagger e^{-i\theta_b} + bb e^{i\theta_b} \right)\right]
\end{eqnarray}
where we have removed the constant term $\sinh^2 u+\sinh^2 v$. 

The 2-mode displacement operator is
\begin{equation}\label{2modesd}
D(\chi)=\exp\left[\chi a^\dagger b- \chi^* a b^\dagger\right]
\end{equation}
and its
action of (\ref{2modesd} on the creation/annihilation operators is 
\begin{eqnarray}\label{a1modesq}
\nonumber D^\dagger(s,\phi)\,a\, D(s,\phi)&=&  a\, \cos s + b e^{i\phi}\sin s\\*
\nonumber D^\dagger(s,\phi)a^\dagger D(s,\phi)&=& a^\dagger\, \cos s + b^\dagger e^{-i\phi}\sin s\\*
\nonumber D^\dagger(s,\phi)\,b\, D(s,\phi)&=&  b\, \cos s - a e^{-i\phi}\sin s\\*
D^\dagger(s,\phi)b^\dagger D(s,\phi)&=& b^\dagger\, \cos s - a^\dagger e^{i\phi}\sin s
\end{eqnarray}
where we have defined $\chi\equiv s e^{i\phi}$.

Computing the effect of the displacement on each of the 6 different operators in (\ref{1step}) we obtain
\begin{eqnarray}\label{resultdisp}
\nonumber D^\dagger(s,\phi)\,a^\dagger a\, D(s,\phi)&=&a^\dagger a\, \cos^2 s + b^\dagger b\,\sin^2 s+ (1/2)\sin 2s \big(a^\dagger b\, e^{i\phi}+b^\dagger a\, e^{-i\phi}\big)\\*
\nonumber D^\dagger(s,\phi)\,b^\dagger b\, D(s,\phi)&=&a^\dagger a\, \sin^2 s + b^\dagger b\,\cos^2 s- (1/2)\sin 2s\, \big(a^\dagger b\, e^{i\phi}+b^\dagger a\, e^{-i\phi}\big)\\*\nonumber
\nonumber D^\dagger(s,\phi)\,a^\dagger a^\dagger\, D(s,\phi)&=&a^\dagger a^\dagger\, \cos^2 s + b^\dagger b^\dagger\,e^{-2i\phi}\sin^2 s+ a^\dagger b^\dagger\, e^{-i\phi}\sin 2s \\*
\nonumber D^\dagger(s,\phi)\,a\, a\, D(s,\phi)&=&a a\, \cos^2 s + bb\,e^{2i\phi}\sin^2 s+ ab\,e^{i\phi}\sin 2s \\*
\nonumber D^\dagger(s,\phi)\,b^\dagger b^\dagger\, D(s,\phi)&=&b^\dagger b^\dagger\, \cos^2 s + a^\dagger a^\dagger\,e^{2i\phi}\sin^2 s- a^\dagger b^\dagger\, e^{i\phi}\sin 2s \\*
 D^\dagger(s,\phi)\,b\, b\, D(s,\phi)&=& bb\, \cos^2 s + a a\,e^{-2i\phi}\sin^2 s- a b\, e^{-i\phi}\sin 2s 
\end{eqnarray}
Next we compute  $H_{1s,2d}=D^\dagger(s,\phi)H_{1s}D(s,\phi)$. Using (\ref{resultdisp}) we find 
\begin{eqnarray}
\nonumber H_{1s,2d}&=&{\omega_a} \left\{\left[a^\dagger a\, \cos^2 s + b^\dagger b\,\sin^2 s+ \frac{1}{2}\sin 2s \left(a^\dagger b\, e^{i\phi}+b^\dagger a\, e^{-i\phi}\right)\right] \cosh 2t_a  \right.\\ 
&&\nonumber +\frac12\sinh 2t_a \left[\left(a^\dagger a^\dagger\, \cos^2 s + b^\dagger b^\dagger\,e^{-2i\phi}\sin^2 s+ a^\dagger b^\dagger\, e^{-i\phi}\sin 2s\right) e^{-i\theta_a}\right. \\
&&\nonumber  \left .\left.
+ \left(a a\, \cos^2 s + bb\,e^{2i\phi}\sin^2 s+ ba\,e^{i\phi}\sin 2s\right) e^{i\theta_a} \right] \right\}   \\*
&&\nonumber+{\omega_b} \left\{\left[a^\dagger a\, \sin^2 s + b^\dagger b\,\cos^2 s- \frac{1}{2}\sin 2s \left(a^\dagger b\, e^{i\phi}+b^\dagger a\, e^{-i\phi}\right)\right]\cosh 2 t_b  \right.\\*
&&\nonumber +\frac12\sinh 2t_b  \left[\left(b^\dagger b^\dagger\, \cos^2 s + a^\dagger a^\dagger\,e^{2i\phi}\sin^2 s- a^\dagger b^\dagger\, e^{i\phi}\sin 2s\right) e^{-i\theta_b} \right. \\
&&\nonumber  \left .\left.
 +\left( bb\, \cos^2 s + a a\,e^{-2i\phi}\sin^2 s- a b\, e^{-i\phi}\sin 2s \right) e^{i\theta_b} \right]\right\}
\end{eqnarray}
Regrouping terms we get
\begin{eqnarray}
\nonumber H_{1s,2d} &=&g_{1} a^\dagger a+g_{2} b^\dagger b+g_{3} a^\dagger b+g^*_{3} b^\dagger a+g_{4} a^\dagger a^\dagger+g^*_{4} a a\\
&&+g_{5} b^\dagger b^\dagger+g^*_{5} bb+g_{6} a^\dagger b^\dagger+g^*_{6} b a
\end{eqnarray}
where
\begin{eqnarray}
\nonumber g_{1} &=&{\omega_a}\,\cos^2 s\,\cosh 2u + {\omega_b}\,\sin^2s\, \cosh 2v,\\*
\nonumber g_{2} &=&{\omega_a}\,\sin^2 s\, \cosh 2u + {\omega_b}\,\cos^2s\, \cosh 2v\\*
\nonumber g_{3} &=&\frac12\sin 2s\,e^{i\phi}\left({\omega_a}\,\cosh 2u - {\omega_b}\cosh 2v\right) \\
\nonumber g_{4} &=&\frac12\left({\omega_a}\,e^{-i\theta_a}\sinh 2u\,\cos^2 s + {\omega_b}\, e^{-i\theta_b}e^{2i\phi}\sinh 2v\,\sin^2 s\right)\\*
\nonumber g_{5} &=&\frac12\left({\omega_a}\,e^{-i\theta_a}e^{-2i\phi}\sinh 2u\,\sin^2 s + {\omega_b}\, e^{-i\theta_b} \sinh 2v\,\cos^2 s\right)\\*
\nonumber g_{6} &=&\frac12\sin 2s\left({\omega_a}\,e^{-i\theta_a}e^{-i\phi}\sinh 2u - {\omega_b}\, e^{-i\theta_b}e^{i\phi} \sinh 2v\right)
\end{eqnarray}

Applying a one mode rotation of the $a$ operators
\[R_{a}=\exp\left(-i\varphi\, {a^\dagger a}\right)\]
we find
\begin{eqnarray}\label{rotarota}
R_a a R^\dagger_a &=& e^{i\varphi}a \qquad R_a a^\dagger R^\dagger_a =e^{-i\varphi}a^\dagger\\ R^\dagger_a a R_a &=& e^{-i\varphi}a \qquad R^\dagger_a a^\dagger R_a =e^{i\varphi}a^\dagger
\end{eqnarray}
yielding 
\begin{eqnarray}\label{total}
\nonumber H_{T}&=&g_{1} a^\dagger a+g_{2} b^\dagger b+e^{i\varphi}g_{3} a^\dagger b+e^{-i\varphi}g^*_{3} b^\dagger a+e^{2i\varphi}g_{4} a^\dagger a^\dagger\\
&& \quad +e^{-2i\varphi}g^*_{4} a a+g_{5} b^\dagger b^\dagger+g^*_{5} bb+e^{i\varphi}g_{6} a^\dagger b^\dagger+e^{-i\varphi}g^*_{6} b a
\end{eqnarray}
for the resultant  Hamiltonian $H_{T}=R^\dagger_a\, H_{1s,2d}\,  R_a$.

Next we demand two conditions in order to reproduce the interaction Hamiltonian  (\ref{hamo}). First we remove the squeezing terms $a^\dagger a^\dagger$ of the field Hamiltonian. To do so, we fix
\[ \tan^2 s=\frac{{\omega_a}\,\sinh 2u}{{\omega_b}\,\sinh 2v}\]
implying
\begin{eqnarray}
\nonumber g_{1}&=&{\omega_a}\,\cos^2 s\left[\cosh 2u +\frac{\sinh 2u}{\tanh 2v}\right]\\*
\nonumber g_{2}&=&\cos^2 s\left[\frac{{\omega_a}^2}{2{\omega_b}}\, \frac{\sinh 4u}{\sinh 2v}+{\omega_b}\,\cos 2v\right] \\*
\nonumber g_{3}&=&\frac12\sin 2s\,e^{i\phi}\left({\omega_a}\,\cosh 2u - {\omega_b}\cosh 2v\right)\\*
\nonumber g_{4}&=&\frac12{\omega_a}\,\cos^2 s\,\sinh 2u\left(e^{-i\theta_a} +  e^{-i\theta_b}e^{2i\phi}\right)\\*
\nonumber g_{5}&=&\frac12\cos^2s\left(\frac{{\omega_b}^2_a}{{\omega_b}}\frac{\sinh^2 2u}{\sinh 2v}e^{-i\theta_a}e^{-2i\phi} + {\omega_b} \sinh 2v\, e^{-i\theta_b}\right)\\*
\nonumber g_{6}&=&\frac12\sin 2s\left({\omega_a}\,e^{-i\theta_a}e^{-i\phi}\sinh 2u - {\omega_b}\, e^{-i\theta_b}e^{i\phi} \sinh 2v\right)
\end{eqnarray}
Setting  $\theta_b=2\phi +\theta_a-\pi$ yields
\begin{eqnarray}
\nonumber g_{1}&=&{\omega_a}\,\cos^2 s\left[\cosh 2u +\frac{\sinh 2u}{\tanh 2v}\right]\\*
\nonumber g_{2}&=&\cos^2 s\left[\frac{{\omega_a}^2}{2{\omega_b}}\, \frac{\sinh 4u}{\sinh 2v}+{\omega_b}\,\cos 2v\right] \\*
\nonumber g_{3}&=&\frac12\sin 2s\,e^{i\phi}\left({\omega_a}\,\cosh 2u - {\omega_b}\cosh 2v\right)\\*
\nonumber g_{4}&=&0\\*
\nonumber g_{5}&=&\frac12e^{-i(2\phi +\theta_a)}\cos^2s\left(\frac{{\omega_b}^2_a}{{\omega_b}}\frac{\sinh^2 2u}{\sinh 2v}- {\omega_b} \sinh 2v\right)\\*
\nonumber g_{6}&=&\frac12e^{-i(\theta_a+\phi)}\sin 2s\left({\omega_a}\,\sinh 2u + {\omega_b}\, \sinh 2v\right)
\end{eqnarray}
and so  the term corresponding to a squeezing of the field has been eliminated.

To reproduce the interaction part we require $g_3=g_6$, implying
\begin{eqnarray}
e^{i(2\phi+\theta_a)}\left({\omega_a}\,\cosh 2u - {\omega_b}\cosh 2v\right) \nonumber \\
\qquad\qquad =\left({\omega_a}\,\sinh 2u + {\omega_b}\, \sinh 2v\right) \nonumber
\end{eqnarray}
Setting $\theta_a=  2n\pi-2\phi$ gives
\[\frac{{\omega_a}}{{\omega_b}}=\frac{\cosh 2v+\sinh 2v}{\cosh 2u-\sinh 2u}=\frac{e^{2v}}{e^{-2u}}\]
and as a consequence
\[u=\frac{1}{2}\ln\left(\frac{{\omega_a}}{{\omega_b}}\right)-v\]
Finally  we need to demand that
\begin{equation}\label{extrar}
\frac{{\omega_a}}{{\omega_b}}>e^{2v}
\end{equation}
 to  ensure that $u>0$.

Recapitulating, we started from the Hamiltonian $H_0$ and applied two 1-mode squeezing operators, a 1-mode displacement operator and a 1-mode rotation on the field operators
\begin{equation}
\nonumber H_T = R_a^\dagger(\varphi)D^\dagger(s,\phi)S^\dagger_a(u,\theta_a)S^\dagger_b(v,\theta_b) H_0S_a(u,\theta_a)S_b(v,\theta_b)D(s,\phi)R_a(\varphi)
\end{equation}
yielding a Hamiltonian depending on  6 parameters.  By fixing 4 of them
\begin{eqnarray}\label{constr}
s&=&\arctan\sqrt{\frac{{\omega_a}\,\sinh 2u}{{\omega_b}\,\sinh 2v}},\qquad \theta_a=  2n\pi-2\phi\\
\theta_b &=& 2\phi +\theta_a-\pi,\qquad \qquad u=\frac{1}{2}\ln\left(\frac{{\omega_a}}{{\omega_b}}\right)-v 
\end{eqnarray}
with the extra requirement for $v$ given by (\ref{extrar}), we obtain 
the hamiltonian $H_T$
\begin{equation}\label{goodham}
\nonumber H_T =\Omega_a a^\dagger a+\hat\Omega_b b^\dagger b + \hat\lambda (b+b^\dagger)(a^\dagger e^{i(\phi+\varphi)}+ a\, e^{-i(\phi+\varphi)})
 + Z\left(b^\dagger b^\dagger + bb\right)
\end{equation}
where
%
%
%
\begin{eqnarray}\label{pumuki}\nonumber\Omega_a &=&\frac{\sinh 2v\left[\cosh \left[2(C-v)\right] +\frac{\sinh\left[2(C-v)\right]}{\tanh 2v}\right]}{{\omega_a^{-1}}\sinh 2v+\omega_b^{-1}\,\sinh \left[2(C-v)\right]}\\[3mm]
\nonumber\hat\Omega_b&=& \frac{\sinh 2v\left[\omega_a^2\, \frac{\sinh \left[4\left(C-v\right)\right]}{2\sinh 2v}+\omega_b^2\,\cosh 2v\right]}{\omega_b\,\sinh 2v+\omega_a\,\sinh [2(C-v)]}\\[3mm]
\nonumber\hat\lambda&=&\frac{\sqrt{\omega_a\omega_b\,\sinh [2(C-v)]\,\sinh 2v}}{\omega_b\,\sinh 2v+\omega_a\,\sinh [2(C-v)]} \left[\omega_a\,\cosh \left[2(C-v)\right] - \omega_b\cosh 2v\right]\\[3mm]
\nonumber Z&=&\frac12 \frac{\sinh 2v\left(\omega^2_a\frac{\sinh^2 \left[2(C-v)\right]}{\sinh 2v}- \omega_b^2 \sinh 2v\right)}{\omega_b\,\sinh 2v+\omega_a\,\sinh [2(C-v)]}\\[3mm]
\varphi = &kx-\Omega_a t
\end{eqnarray}
with $C=\frac{1}{2}\ln\left(\frac{{\omega_a}}{{\omega_b}}\right)$ and where $2p= {\tanh}^{-1}\big[-2Z/\hat \Omega_b\big]$ .


The rotation is necessary to account for the time evolution on a given trajectory as it is completely decoupled from the rest of parameters. Actually for a particular choice of the displacement parameter phase $\phi$ (for example $\phi=0$)  we trivially get 
\begin{equation}\label{goodham2a}
\hat H_T=\Omega_a\, a^\dagger a+\hat{\Omega}_b\, b^\dagger b + \hat\lambda (b+b^\dagger)(a^\dagger\, e^{i\varphi}+ a\, e^{-i\varphi})+ Z\left(b^\dagger b^\dagger + bb\right)
\end{equation}
Applying another squeezing operator  $S_{b}(p)$ (where $p$ is real) yields
\begin{eqnarray}
\nonumber S^\dagger_a S^\dagger_bb^\dagger b S_b S_a &=&b^\dagger b \, \cosh 2p + \frac12\sinh 2p\left(b^\dagger b^\dagger + bb \right)+  \sinh^2 p\\
\nonumber S^\dagger_a S^\dagger_b bb S_b S_a &= &bb \cosh^2 p+b^\dagger b^\dagger\,\sinh^2 p +  b^\dagger b\, \sinh 2p +\frac12\sinh 2p \\
S^\dagger_a S^\dagger_bb^\dagger b^\dagger S_b S_a &=& b^\dagger b^\dagger \cosh^2 p+bb\,\sinh^2 p + b^\dagger b\, \sinh 2p +\frac12\sinh 2p 
\end{eqnarray}
and so the interaction Hamiltonian $H_T=S^\dagger_b(p)\hat H_T S_b(p)$, after eliminating constant terms, is
\begin{eqnarray}
\nonumber H_T&=&\Omega_a a^\dagger a+\hat\Omega_b\Big[b^\dagger b  \cosh 2p + \frac12\sinh 2p\left(b^\dagger b^\dagger + b b \right) + \sinh^2 p\Big]\\
&&\nonumber  + \hat\lambda (\sinh q + \cosh q) (b+b^\dagger)(a^\dagger e^{i\varphi}+ a e^{-i\varphi})\\
&&\nonumber+ Z\big(f b \cosh^2 p+b^\dagger b^\dagger\sinh^2 p +  b^\dagger b\sinh 2p \\
&&\qquad +b^\dagger b^\dagger \cosh^2 p+bb\,\sinh^2 p + b^\dagger b\, \sinh 2p\big)
\end{eqnarray}
which can be rewritten as
\begin{eqnarray}
\nonumber H_T&=&\Omega_a\, a^\dagger a+\left(\hat\Omega_b \cosh 2p + 2Z \sinh 2p\right)b^\dagger b\\ &&+ e^{q}\hat\lambda  (b+b^\dagger)(a^\dagger\, e^{i\varphi}+ a\, e^{-i\varphi})\\
&&+(b^\dagger b^\dagger + bb)\Big(Z\cosh 2p + \frac{\hat\omega}{2}\sinh 2p\Big)
\end{eqnarray}

Fixing a value of $p$ such that
\[2p=\tanh^{-1}\left(\frac{-2Z}{\hat \omega}\right)\]
yields the Hamiltonian  
\begin{eqnarray}
\nonumber H_T&=&\Omega_a a^\dagger a\!+\!\sqrt{\hat\Omega_b^2\!-4Z^2}\,b^\dagger b + e^{q}\hat\lambda  (b+b^\dagger)(a^\dagger e^{i\varphi}\!+\! a e^{-i\varphi})
\end{eqnarray}
We can rewrite this as an Unruh DeWitt hamiltonian
\begin{equation}
\nonumber H_T  = \Omega_a a^\dagger a+{\Omega_b}\,b^\dagger b + \lambda  (b+b^\dagger)(a^\dagger e^{i\varphi}+ a e^{-i\varphi})
\end{equation}
where
\begin{equation}\label{res1}
\lambda =e^{p}\hat \lambda\qquad\Omega_b = \sqrt{\hat\Omega_b^2-4Z^2}
\end{equation}


\bibliographystyle{ieeetr}       
\bibliography{references}
\end{document}